\documentclass[12pt]{article}
\pdfoutput=1
\usepackage{amsmath}
\usepackage{amssymb}
\usepackage{amsfonts}
\usepackage{bbold}
\usepackage{graphicx}
\usepackage{axodraw}
\usepackage{verbatim}
\usepackage{slashed}
\usepackage{array}

\def\eq#1{{eq.~(\ref{#1})}}

\def\vev#1{\left\langle #1\right\rangle}

\def\Im{\mbox{Im}\,}

\def\hbar{\hspace{0pt}\raisebox{1pt}{$-$} \hspace{-7pt} h}

\def\c{\hspace{-5pt}}

\newcommand{\be}{\begin{equation}}
\newcommand{\ee}{\end{equation}}
\newcommand{\bea}{\begin{eqnarray}}
\newcommand{\eea}{\end{eqnarray}}
\newcommand{\nn}{\nonumber}

\newcommand{\comb}[1]{{\color{blue} \sffamily #1}} 

\def\5{\overline 5}

\renewcommand{\Im}{\mbox{Im}}

\let\vev\VEV

\def\roughly#1{\mathrel{\raise.3ex\hbox{$#1$\kern-.75em
      \lower1ex\hbox{$\sim$}}}} 

\def\e6{E(6)}
\def\321{$SU(3)_{c}\otimes SU(2)_L \otimes U(1)$}
\def\10{SO(10)}

\def\422{SU(4) $\otimes$ SU(2) $\otimes$ SU(2)}

\newcommand{\beq}{\begin{equation}}
\newcommand{\eeq}{\end{equation}}
\newcommand{\bac}{\beq\begin{array}}
\newcommand{\eac}{\end{array}\eeq}
\newcommand{\ba}{\begin{array}}
\newcommand{\ea}{\end{array}}
\newcommand{\esp}{\end{split}}
\newcommand{\bsp}{\begin{split}}

\newcommand{\bmat}{\begin{pmatrix}}
\newcommand{\emat}{\end{pmatrix}}
\def\author#1{\begin{center} #1\end{center}}

\begin{document}

\textwidth 17cm
\textheight 22cm
\hoffset-1.5cm
\voffset-1cm

\begin{titlepage}
  \newcommand{\AddrSISSA}{{\sl \small SISSA and INFN, Sezione
    di Trieste,\\ Via Bonomea 265, 34136 Trieste, Italy}}
  \newcommand{\AddrLiege}{{\sl \small IFPA, Dep. AGO, 
      Universite de Liege, Bat B5,\\ \small \sl Sart
      Tilman B-4000 Liege 1, Belgium}}
\vspace*{1.5cm}
\begin{center}
  \textbf{\Large Leptogenesis in the presence of}\\[2mm]
  \textbf{\Large exact flavor symmetries}
  \\[10mm]
  D. Aristizabal Sierra$^{a,}$\footnote{e-mail address: daristizabal@ulg.ac.be},
  Federica Bazzocchi$^{b,}$\footnote{e-mail address: fbazzo@sissa.it}
  \vspace{0.8cm}
  \\
  $^a$\AddrLiege.\vspace{0.4cm}\\
  $^b$\AddrSISSA.\vspace{0.4cm}\\
\end{center}
\vspace*{0.5cm}
\begin{abstract}
  In models with flavor symmetries in the leptonic sector leptogenesis
  can take place in a very different way compared to the standard
  leptogenesis scenario. We study the generation of a $B-L$ asymmetry
  in these kind of models in the flavor symmetric phase pointing out
  that successful leptogenesis requires (i) the right-handed neutrinos
  to lie in different irreducible representations of the flavor group;
  (ii) the flavons to be lighter at least that one of the right-handed
  neutrino representations. When these conditions are satisfied
  leptogenesis proceeds due to new contributions to the CP violating
  asymmetry and---depending on the specific model---in several
  stages. We demonstrate the validity of these arguments by studying
  in detail the generation of the $B-L$ asymmetry in a scenario of a
  concrete $A_4$ flavor model realization.
\end{abstract}
\end{titlepage}
\setcounter{footnote}{0}

%
%

\section{Motivation}
\label{sec:motivation}
Observational data from the abundances of light elements (D, $^3$He,
$^4$He and Li) in addition to precision observations of the cosmic
microwave background (CMB) temperature fluctuations allow the
determination of the cosmic baryon asymmetry,
$Y_{\Delta_B}=(n_B-n_{\bar B})/s=(8.75 \pm 0.23)\times 10^{-11}$ (with
$n_B$ ($n_{\bar B}$) the baryon (antibaryon) number density and $s$
the entropy density) \cite{Hinshaw:2008kr}. Though the conditions for
dynamically generating this quantity are well known and established
\cite{Sakharov:1967dj} the cosmic baryon asymmetry poses a puzzle in
particle physics: the standard model (SM) fails to explain such a
large asymmetry, thus implying the presence of new physics accounting
for $Y_{\Delta_B}$.

Leptogenesis is a scenario in which a lepton asymmetry $Y_{\Delta_L}$
is generated in the lepton sector and partially reprocessed into
$Y_{\Delta_B}$ by SM electroweak sphaleron processes (for a
comprehensive review see \cite{Davidson:2008bu}). The generation of
$Y_{\Delta_L}$ requires, in addition to CP violation and departure
from thermodynamical equilibrium, lepton number breaking. Accordingly,
in these class of scenarios two in principle unrelated puzzles are
linked, the origin of neutrino masses and the baryon asymmetry. Among
all the possible neutrino mass models present in the literature the
standard seesaw (type I seesaw) \cite{seesaw} provides the framework
for {\it standard leptogenesis}, in which the lepton asymmetry
proceeds via the out-of-equilibrium and CP violating decays of the
lightest right-handed (RH) electroweak singlet neutrino.

Most of the studies of leptogenesis are based on the assumption that
there is no new physics between the lepton number breaking scale and
the electroweak scale that can sizable affect the way in which
leptogenesis takes place. Though some analysis in scenarios including
flavor symmetries above the electroweak scale have been done, and have
proved that the presence of new energy scales and new degress of
freedom may have an impact on the way leptogenesis proceeds
\cite{Hagedorn:2009jy,Bertuzzo:2009im,
  AristizabalSierra:2009ex,Felipe:2009rr}, all of them are based on
the same assumption, namely the lepton number breaking scale is below
the scale at which the flavor symmetry is broken \footnote{The
  exceptions being references \cite{AristizabalSierra:2007ur,
    AristizabalSierra:2009bh}.}.

The idea to ascribe to a flavor symmetry to explain particle masses
and mixings dates backs to the late 1970's
\cite{Froggatt:1978nt}. Originally flavor symmetries were introduced
to explain quark structures and only after neutrino oscillation data
the use of horizontal symmetries in the lepton sector has become more
challenging and interesting. In particular, in the last years it has
been shown that lepton mixing may be well described by discrete non
Abelian symmetries (see \cite{Altarelli:2010gt} and references therein
for further details).  Given that in these kind of models both, the
lepton number and flavor breaking scales are free parameters the
question about how does leptogenesis proceeds in the flavor symmetric
phase proves to be quite reasonable.

In more detail, let us suppose to have a non Abelian flavor symmetry
group $G_F$ under which the RH neutrinos and left-handed SM leptons
have definitive transformations. We introduce a cutoff scale $\Lambda$
since we will deal with non-renomalizable operators.  The scenario
which we are interested in is the following: the lepton number
breaking scale, characterised by the RH neutrino mass, $M_N$, is
larger than the scale at which $G_F$ is broken, $v_F$. This means that
the Yukawa mass matrices above and below $v_F$ are different: in
particular the Yukawa Dirac matrix below $v_F$ is proportional to the
Dirac mass matrix---the proportionality factor represented by the SM
Higgs vacuum expectation value (vev). Flavon masses, $M_\phi$, are
taken as free parameters, clearly not too far from $v_F$ but above
it. Clearly it holds $\Lambda> M_N,M_\phi> v_F$.

Our discussion is based on the class of symmetries that explain
neutrino masses and lepton mixings and may be be generalized to any
flavor symmetry, Abelian or not, discrete or continuos.  We will
exemplify our arguments by doing a full analysis of the generation of
the $B-L$ in a concrete model that at low scale exhibits exact
TriBiMaximal (TBM) mixing at leading order. The reason for this choice
is very simple: it has been shown that when exact TBM mixing is
induced by type I seesaw the CP violating asymmetry is zero and
acquires a non-vanishing value only when lepton mixing deviates from
TBM \cite{Jenkins:2008rb,Riazuddin:2008th,AristizabalSierra:2009ex}.
In type II seesaw models this could be also the case if only two
electroweak triplets are present and they both are family singlets
\cite{deMedeirosVarzielas:2011tp}. Extended models featuring interplay
between type I and II seesaws have been also analysed and the
conclusion is that in these cases it is also possible to identify a
generic class of minimal models in which the CP asymmetry vanishes as
well \cite{diego-federica-ivo}.

The most recent analysis indicates that TBM is not anymore in perfect
agreement with neutrino experimental data since the reactor angle
predicted by TBM is zero, while this value is now excluded at
$3\sigma$ level \cite{Schwetz:2011zk}. However TBM remains a good
approximation for the lepton mixing matrix and we will consider a
model that predicts exact TBM for its simplicity in showing the
feasibility of leptogenesis in the regime $\Lambda>M_N,M_\phi>v_F$.

The paper is organized in the following way: next section is general
and we enumerate the general conditions necessary to obtain a CP
asymmetry, $\epsilon_N$, in the flavor symmetric regimen. Section
\ref{setup} shows how the proposed conditions work by mean of a
complete analysis of a specific model based on the flavor symmetry
$A_4$. The model main features and neutrino phenomenology are briefly
discussed and the generation of the $B-L$ asymmetry is explained in
detail.  Section \ref{sec:conclusions} is devoted to our
conclusions. The calculation of the reaction densities necessary for
the analysis of the washout processes studied in sec. \ref{setup} are
given in appendix \ref{sec:app-washout}.
\section{Leptogenesis in the flavor symmetric phase}
\label{sec:fel}
We have already anticipated in the introduction that in models for
leptonic flavor mixing four energy scales can be distinguished, namely
a cutoff scale $\Lambda$---or in general a scale of heavy matter---,
the lepton number breaking scale---determined by the RH neutrino
masses---$M_N$, the flavons scale $M_\phi$---determined by the masses
of the scalars that trigger the flavor symmetry breaking---and the
scale at which the flavor symmetry is broken, $v_F$. Though $\Lambda>
M_N, M_\phi, v_F$ the scales $M_N$, $M_\phi$ and $v_F$, being free
parameters, can follow any hierarchy. Since we are concerned about
leptogenesis in the flavor symmetric phase it is clear $v_F < M_N,
M_\phi$. This constraint in turn has an implication: if the flavor
symmetry enforces the RH neutrinos to belong to the same $G_F$ irreducible 
representation, $R$, leptogenesis will not be achievable: in the
flavor symmetric phase the RH neutrinos have a common universal mass
and therefore the CP violating asymmetry $\epsilon_N$ vanishes
\cite{Branco:2006hz}.

As a consequence, in these kind of models viable leptogenesis requires
RH neutrinos to belong to different $G_F$ irreducible representations $R_i$, so a
mass splitting among the masses of the different representations can
be accomodated. Let us assume the existence of $k$ electroweak lepton
doublets placed in $r$ representations $L_r$, $m$ RH neutrinos lying
in $p$ representations $N_p$ and $n$ electroweak singlet scalars
arranged in $q$ representations $S_q$. Assuming the SM Higgs $SU(2)$
doublet, $H$, to be a $G_F$ singlet the $i$-th RH neutrino
representation can only decay to final states containing
$L_i$. Accordingly, three type of models can be distinguished:
\begin{enumerate}
\item \label{model-case1} For any RH neutrino representation $N_i\sim R_i$ there is
  a lepton doublet representation $\bar{L}_i \sim R'_i$ with which a gauge flavor invariant
  renormalizable operator $\bar{L}_i\,N_i\,H$ can be built
  \footnote{Here $i$ labels the index representation, not the flavor
    index and with $N_i\sim R_i (\bar{L}_i \sim R'_i)$ we indicate that $N_i(\bar{L}_i)$ transforms as the representation $R_i(R'_i)$. We introduce the generic $R'_i$ representation for $\bar{L}_i$ to be as general as possible, defining $R'_i$ as the representation that contracted with $R_i$ has in its Clebsch-Gordan series  a singlet of $G_F$. Indeed for the discrete groups  with real triplet or doublet representation it holds that $R'_i=R_i$.  For these kind of groups it makes sense defining the $R_i^*$  only when  the physical field--a scalar or a fermion--is complex.}.
\item \label{model-case2} Only for a set of the RH neutrino
  representations a gauge flavor invariant renormalizable operator
  $\bar L_i\,N_i\,H$ exist
\item \label{model-case3} For non of the RH neutrino representations
  the operator $\bar L_i\,N_i\,H$ can be built.
\end{enumerate}
In cases \ref{model-case1} and \ref{model-case2} the standard one-loop
vertex and wave-function corrections to the tree-level decay exist
however the CP violating asymmetry derived from their interferences
and the corresponding tree-level process vanishes. The proof of this
statement is easy. Suppose $N_i$ transforms as the $R_i$
representation of $G_F$. To recover the correct kinetic term we know
$R_i^* R_i=\delta_{\alpha_i\beta_i}$ should hold (here
$\alpha_i,\beta_i$ are the indices of the $R_i$ representation). The same applies for $\bar{L}_i\sim R'_i$. Now,
the standard contribution to the CP asymmetry involves the imaginary
part of the off-diagonal elements of the matrix $Y\, Y^\dagger$, where
$Y$ is a generic Yukawa coupling matrix defined in the stage of
unbroken flavor symmetry. This Yukawa coupling matrix is given by the
Clebsch-Gordan coefficients arising from the contractions $R'_i\,R_i$,
thus implying that the matrix $Y\, Y^\dagger$ is determined in turn by
the contractions $R'_i\,R_i\,R_i^*\,{R'}_i^*$, which demonstrates that $Y
Y^\dagger$ is diagonal.
Viable leptogenesis is, therefore, possible only if new contributions
($\epsilon_N^\text{New}$) to $\epsilon_N$ are present, and this is
possible only if at least for one of the RH neutrino representations
the condition $M_N>M_\phi$ is satisfied \footnote{This condition, in
  addition with the mass splitting among the different RH neutrino
  representations, guarantee that the novel loop corrections to the
  tree-level decay contain an imaginary part, an essential requirement
  for $\epsilon_N^\text{New}\neq 0$.}

In case \ref{model-case3} it is possible that $N_i$ may have $n>2$
body decays by means of non renormalizable operators. In this scenario
one should modify the standard case in order to include $n$-body
decays. However, due to space suppression factors the CP asymmetry
generated is expected to be small. For this reason in what follows we
do not consider this case.

Assuming the flavons are lepton conserving states no $B-L$ asymmetry
can be generated via $\phi$ decays. However, once the condition
$M_N>M_\phi$ is satisfied they can play an essential role in the
generation of the $B-L$ asymmetry, not only because they lead to novel
contributions to $\epsilon_N$, but because in some cases they can even
allow some RH neutrino representations to have new decay modes that
can change the way in which leptogenesis takes place.  Once the
conditions discussed above are satisfied not much more, from a general
perspective, can be said and the way in which the $B-L$ asymmetry is
generated depends upon the particular flavor model. Hereafter are
discussion will rely on a particular $A_4$ flavor model realization.
\section{Setup}
\label{setup}
We consider the non-supersymmetric version of a model inspired by the
Altarelli-Feruglio model discussed in \cite{Altarelli:2006kg} of which
the type-I seesaw formulation has been analyzed in
\cite{Bazzocchi:2009da}. In the original model supersymmetry is
introduced to induce the correct spontaneous breaking of the flavor
symmetry. Here we assume that by adding additional discrete Abelian
symmetries or \emph{ad hoc} soft terms the scalar potential may be
arranged in such a way that the desired breaking is realized. At this
level the model presented is still a toy model, however our findings
will hold even in its supersymmetric version.

\subsection{The $\pmb{A_4}$ group}
\label{sec:a4-group}
Before entering into the details of the model, for completeness, we
will briefly discuss the basic ingredients of the $A_4$ discrete group
in which the model presented here is based. $A_4$ is the group of even
permutations of 4 objects. It has 4 irreducible representations: one
triplet and three singlets $1,1',1''$. $A_4$ may be thought as
generated by two elements $S,T$ satisfying 
\beq 
S^2=T^3=(ST)^3=1\,.
\eeq 

 
In what follows we will work in the $A_4$ basis in which the triplet representation of $T$ is diagonal, namely
 \bea
 S=\frac{1}{3}
 \left(
   \begin{array}{ccc} 
     -1 & 2 & 2\\
     2 & -1 & 2\\
     2 & 2 & -1
   \end{array} 
 \right)\,,&\quad&  
 T=\left(
   \begin{array}{ccc}
     1&0&0\\
     0&\omega&0\\
     0&0& \omega^2
   \end{array}
 \right)
\eea
 with $\omega^3=1$. The multiplication rules in this basis are given by
 \bea
 && (a b)_1= ( a_1 b_1+ a_2 b_3+ a_3 b_2)\,,\nn\\
 &&  (a b)_{1'}= ( a_3 b_3+ a_1 b_2+ a_2 b_1) \,,\nn\\
 &&  (a b)_{1''}= ( a_2 b_2+ a_1 b_3+ a_3 b_1) \,,\nn\\
 &&   (a b)_{3_s}= ( 2 a_1 b_1- a_2 b_3- a_3 b_2,2 a_3 b_3- a_2 b_1- a_1 b_2,2 a_2 b_2- a_3 b_1- a_1 b_3)\,,\nn\\
 &&    (a b)_{3_a}= ( a_2 b_3- a_3 b_2,a_1 b_2- a_2 b_1, a_3 b_1- a_1 b_3)\,,\nn\\
 \eea
 where $a$ and $b$ are triplets of $A_4$, namely $a\sim (a_1,a_2,a_3),b\sim (b_1,b_2,b_3)$ and for the singlet representations the multiplication rules are trivial
 \beq
 1' \otimes 1''  =1\,,\quad 1' \otimes 1=1' \,,\quad 1'' \otimes 1=1''\,.
 \eeq

\subsection{The model}
\label{sec:the-model}  
In our model four RH neutrinos are added to the SM field
content. Three of them, $\nu_T$, form an $A_4$ triplet while the
fourth, $\nu_4$, is an $A_4$ singlet. The SM lepton doublets,
$l_1,l_2,l_3$, transform as an $A_4$ triplet. For simplicity we assign
flavor charges using the Weyl spinor notation. The RH 4-dim Majorana
fermion will be therefore defined as 
\beq N=\left(
  \begin{array}{c}
    (\nu_R)^C \\ 
    \nu_R\end{array}
\right)\,.
\eeq

It is important to notice   that if $l\sim(l_1,l_2,l_3)$ transforms as a  triplet the requirement of recovering the correct kinetic term according to the group multiplication rules imposes that ${l}^\dag$ transforms as a triplet but  ordered as ${l}^\dag\sim({l}^\dag_1,{l}^\dag_3,{l}^\dag_2)$. On the contrary  we order $\nu_T^\dag$ as  $(\nu^\dag_1,\nu^\dag_2,\nu^\dag_3)$ thus $\nu_T\sim(\nu_1,\nu_3,\nu_2)$. 
 
RH charged  leptons transform as the 3 one dimensional representation of $A_4$,  namely $1,1',1''$.  Two $A_4$ scalar triplets, $\phi_T$ and $\phi_S$, our flavons,  are added.  Once the flavor symmetry is broken they will give rise to the correct  mass matrices.

\subsubsection{ Neutrino mass matrices}
\label{sec:neu-mm}
Given the field content previously described the Lagrangian for the
lepton sector---not including the kinetic terms---reads as
 \bea
 \label{lag}
 -\mathcal{L}&=& \frac{M_{\nu T}}{2} (\nu_{TR}^{\dag} \nu^c_{TR} + \nu_{TR}^{c\dag} \nu_{TR})_1+   \frac{M_{\nu4}}{2} (\nu_{4R}^{c\dag} \nu_{4R}+ \nu_{4R}^\dag \nu_{4R}^c)_1 \nn\\
 & +&\lambda   [\nu_{TR}^{\dag}\nu^c_{TR} \phi_S ]_1 +  \lambda^* [\nu_{TR}^{c\dag} \nu_{TR} \phi^*_S ]_1 \nn\\
  & +& \xi  [\nu_{TR}^{\dag}\phi_S ]_1\nu^c_{4R} +  \xi^* \nu_{4R}^{c\dag}[ \nu_{TR} \phi^*_S]_1 \nn\\
 &+& y_1 \epsilon_{\alpha \beta}(\nu_{TR}^\dag l_L^\alpha )_1H^\beta + y^*_1 \epsilon_{\alpha \beta}( l_L^{\dag\alpha}\nu_{TR} )_1H^\beta\nn\\
 &+&y_2 \frac{1}{\Lambda}  \epsilon_{\alpha \beta}[\nu_{TR}^\dag l_L^\alpha \phi_S]_1H^\beta+y_2^*  \frac{1}{\Lambda}  \epsilon_{\alpha \beta}[l_L^{\dag\alpha}\nu_{TR} \phi^*_S ]_1H^{*\beta} \nn\\
 &+&y_3 \frac{1}{\Lambda}  \epsilon_{\alpha \beta}\nu_{4R}^\dag[ l_L^\alpha \phi_S]_1H^\beta+  y_3^*\frac{1}{\Lambda}  \epsilon_{\alpha \beta}[l_L^{\dag\alpha} \phi^*_S]_1\nu_{4R} H^\beta \nn\\
 &+&y_e^i \frac{1}{\Lambda} e^c_{Ri} [ \phi_T l_L^\alpha]_i
 \tilde{H}^\alpha+ \text{H.c.}\,.
\eea 
We have assumed the presence of an Abelian $Z_N$ with $N>2$ that
forbids $\phi_{S,T}$ and $\phi_{S,T}^*$ to have the same couplings and
we are assuming the flavons to be complex fields.  Note that in this
class of models this kind of Abelian symmetries are always present to
prevent interferences between $\phi_T$ and $\phi_S$.

Note also that we can rephase the RH neutrinos to make $M_{\nu T}$ and
$M_{\nu 4}$ real. After doing so we still have the freedom to rephase
the lepton $A_4$ triplet and the flavon $A_4$ singlet, absorbing in
that way other two CP phases. These basis rotations allow us to choose
$\lambda$ and $y_3$ to be real.

In \eq{lag} the last row describes charged leptons: $i$ stays for
$1,1',1''$ and the $[...]_i$ stays for the triplet contractions in the
one dimensional representations. In the other rows $[...]_1$ stays for
2 or 3 triplets contracted in a singlet. Greek indices $\alpha$ and
$\beta$ label $SU(2)_L$ degrees of freedom, $H$ the SM Higgs doublet,
and as usual $\tilde{H}= i \sigma_2 H$. We assume that the additional
$Z_N$ symmetry forbids the coupling of $\phi_T$ and $\phi_S$ to
neutrinos and charged leptons respectively. The $A_4$ basis chosen is
useful because when $\phi_T$ develops a vev according to
$\vev{\phi_T}\sim v_T (1,0,0)$ the charged lepton Yukawa mass matrix
is diagonal. On the other hand when $\phi_S$ acquires the vev
$\vev{\phi_S}\sim v_S(1,1,1)$ the Dirac Yukawa matrix, $Y_\nu$, and
the RH neutrino mass matrix, $\mathcal{M}_N$ are diagonalized by the
so-called TBM mixing matrix. Thus after electroweak symmetry breaking
the light neutrino mass matrix,

\beq
m_\nu \sim - m_D^T\cdot \mathcal{M}_N^{-1}\cdot m_D
\eeq
is diagonalized by the TBM mixing matrix as well. Clearly $m_D = Y_\nu v_H$, with $\vev{H}=v_H$.

Both $\mathcal{M}_N$ and $Y_\nu$ get a contribution above ($>$) and below ($<$) the scale $v_F \sim v_T\sim v_S$, so we may write
\bea
\mathcal{M}_N&=& \mathcal{M}^>_N+ \mathcal{M}^<_N\,,\nn\\
Y_\nu&=& Y_\nu^> +Y_\nu^<\,,
\eea
with
\bea
\label{matrices}
 \mathcal{M}^>_N &\sim&  \left( \begin{array}{cccc}  A&0&0&0\\0&0&A&0\\ 0&A&0&0\\ 0&0&0&D\end{array} \right)\,,\quad
 Y_\nu^>\sim \left( \begin{array}{ccc}  a&0&0\\0&0&a\\ 0&a&0\\ 0&0&0\end{array} \right) \nn \\
  \mathcal{M}^<_N &\sim&  \left( \begin{array}{cccc}  2 B&-B&-B&C\\-B&2 B&-B&C\\ -B&-B&2B&C\\C&C&C &0\end{array} \right)\,,\quad
  Y_\nu^<\sim\left( \begin{array}{ccc}  2 b&-b&-b\\-b&2 b&-b\\ -b&-b&2b\\c&c&c \end{array} \right)\,.
\eea
with $A\sim M_{\nu T},B \sim\lambda v_S,C\sim \xi v_S$, $D\sim M_{\nu 4}$
and $a\sim y_1,b\sim y_2 v_S/\Lambda,c\sim y_3 v_S/\Lambda$ for the RH
neutrino and Yukawa Dirac mass matrix respectively. Without loss of
generality $v_{S}$ may be taken real.

Defining $\epsilon_S= v_S/\Lambda$, clearly in the limit
$\epsilon_S\to0$ the symmetry is restored and the light neutrinos are
degenerate. Thus we expect in the majority of the cases a
quasi-degenerate (QD) neutrino mass spectrum in which $|y_1|^2/M_{\nu T}$
controls the absolute mass scale while $y_2 , \lambda,
\epsilon_S=v_S/\Lambda$ parametrize the neutrino atmospheric mass
splitting. We may find an approximate analytical solution for the
spectrum expanding in $\epsilon_S$. We have
\bea
\label{spect}
 m_0&\sim &|y_1|^2 \frac{v_H^2}{M_{\nu T}}\,,\nn\\
\frac{\Delta m^2_{sol}}{\Delta m^2_{atm}}&\sim &\frac{1}{2}+ \frac{\epsilon_S}{2} (\epsilon_X +\epsilon_X^2)\frac{(|y_1|^2 \xi^2 -2 |y_1| y_3 \xi \cos \phi_{y_1} +y_3^2 \cos 2 \phi_{y_1})}{|y_1| (|y_1| \lambda -2 |y_2| \cos \Delta{\phi_{12}})}\,,
\eea
with $\epsilon_X=(M_{\nu T}-M_{\nu4})/ M_{\nu T} $, $ \phi_{y_{1,2}} =\text{Arg}(y_{1,2})$ and $\Delta{\phi_{12}}= \phi_{y_{1}} - \phi_{y_{2}} $.

\begin{figure}
  \centering
  \includegraphics[width=13cm,height=8cm]{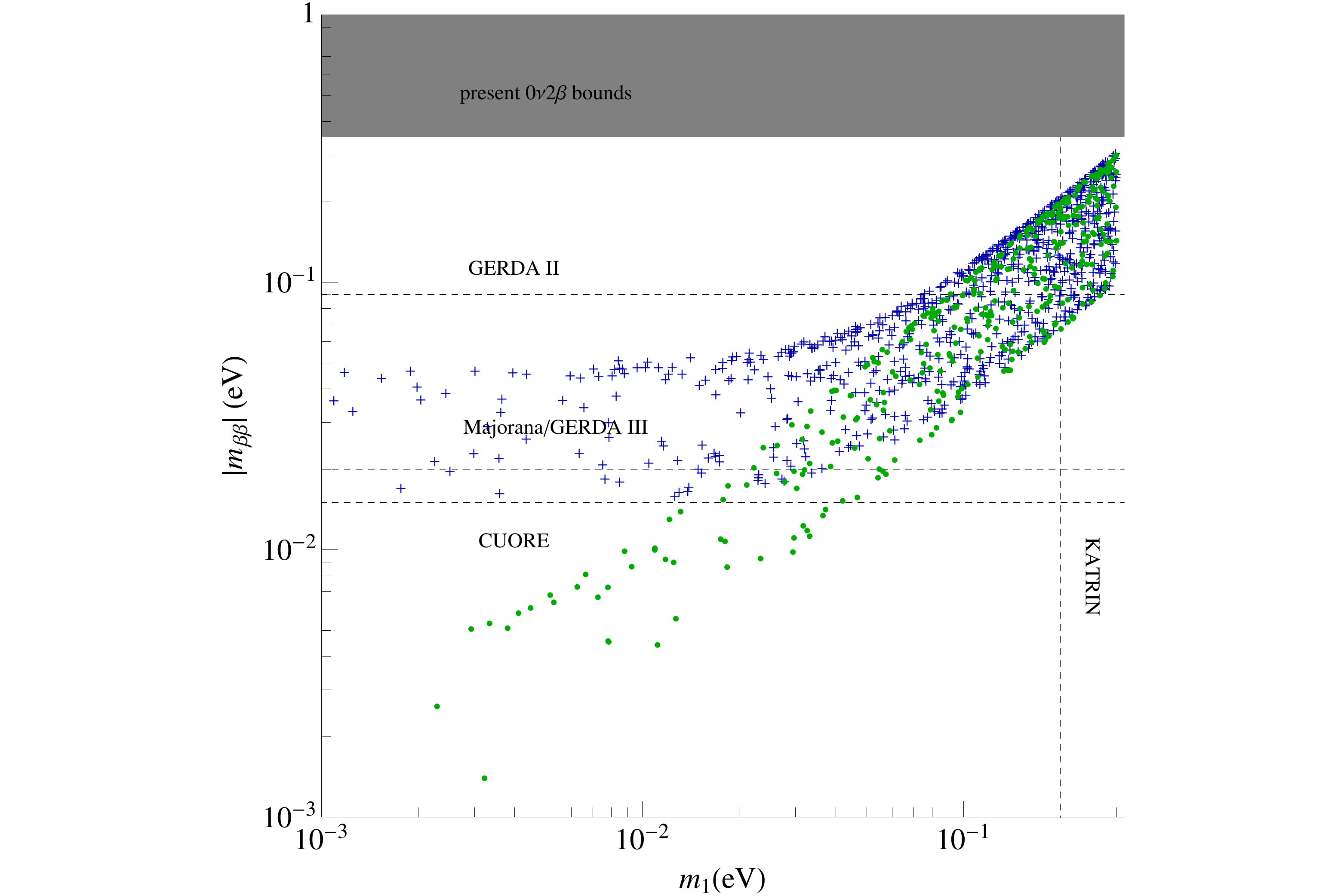}
  \caption{\it The predictions for $|m_{\beta\beta}|$ as a function of
    the lightest neutrino mass.  The natural spectrum predicted by the
    model is QD, with both normal ordering (NO) and inverse ordering
    (IO) as indicated by the analytical approximations. Both QD-NO
    (QD-IO) and NH (IH) spectrum are indicated with green points (blue
    crosses). See the text for more details.}
  \label{fig:mbb}
\end{figure}

Equation \eqref{spect} holds only in the regime
$\epsilon_S<\epsilon_X$. When $\epsilon_X <\epsilon_S$ the analytical
expressions become more cumbersome because the QD scenario is broken
and both, the normal hierarchical (NH) and inverse hierarchical (IH)
neutrino mass spectra become possible. The spectrum predicted by our
model is shown in fig.\ref{fig:mbb} by means of $m_{\beta\beta}$, the
parameter relevant for neutrinoless double beta decay defined as
$m_{\beta \beta} = [U_{\text{TBM}}\, \text{diag}(m_1, \, m_2, \,
m_3) \, U_{\text{TBM}}^T ]_{11}$.  It has been obtained by numerically
diagonalizing the neutrino mass matrix below the scale $v_S$, fixing
the neutrino mixing angles according to the TBM scheme and requiring
the solar and atmospheric mass splittings to lie within their currents
$3\,\sigma$ ranges \cite{Schwetz:2011zk}. The parameters entering in
the mass matrix were varied according to: $[10^{-3},10^{-1}]$ for the
dimensionless parameters, $\epsilon_S$ constrained to be below
$10^{-1}$ and RH neutrino masses in the range $[10^7,10^{14}]$
GeV. For all the points $\Lambda$ was taken to be larger than the
heaviest RH neutrino mass. For completeness in the figure we have
shown the future experimental bounds on $|m_{\beta\beta}|$ and $m_1$.

For what concerns the  flavon sector  the $Z_N$ symmetries allows only the  mass term
 \beq
 M_\phi^2 (\phi_S \phi_S^*)_1\,.
 \eeq
 
 Recalling now that $\phi^*_S\sim(\phi^*_{S1},\phi^*_{S3},\phi^*_{S2})$ we have that the flavon
 mass matrix is diagonal and CP even and odd states are degenerate.

\subsubsection{Flavon Interaction Matrices}

Above $v_F$ instead of  $ \mathcal{M}^<_N , Y_\nu^<$ we have flavon-neutrino and flavon-Higgs-neutrino interaction matrices $\mathcal{I}_{N ,\nu}$. Starting from the interactions
\bea
 \label{lagI}
 -\mathcal{L}_I&=& 
 \frac{1}{2} \lambda [\nu_{TR}^{\dag}\nu_{TR}^c \phi_S ]_1 +\frac{1}{2}  \lambda^* [\nu_{TR}^{c\dag} \nu_{TR} \phi^*_S]_1 \nn
  + \xi  [\nu_{TR}^{\dag} \phi_S ]_1\nu^c_{4R} + \xi^*  \nu_{4R}^{c\dag}[ \nu_{TR} \phi^*_S  ]_1 \nn\\
  &+& \frac{1}{\Lambda}y_2  \epsilon_{\alpha \beta}[\nu_{TR}^\dag l_L^\alpha \phi_S]_1H^\beta+  \frac{1}{\Lambda} y_2^* \epsilon_{\alpha \beta}[l_L^{\dag\alpha}\nu_{TR} \phi^*_S]_1H^{*\beta}\,,
   \eea
we may write them using the 4-component spinors $N$ and $P_L l=l_L$ as
\bea
\label{4comp}
 && \frac{1}{2}\mathcal{I}^k_{N_R }(\phi^*_{S_k})_{ij} \bar{N}_i P_R N_j +  \frac{1}{2}\mathcal{I}^k_{N_L }(\phi_{S_k})_{ij} \bar{N}_i P_L N_j \nn\\
&&+   \mathcal{I}_{D_L }^k(\phi_{S_k})_{ij}  \epsilon_{\alpha \beta}\bar{N}_i P_L  l^\alpha_j H^\beta+  \mathcal{I}_{D_R }^{k}(\phi_{S_k}^*)_{ij}  \epsilon_{\alpha \beta}\bar{l}^\alpha_i P_R N_j \tilde{H}^\beta\,,
\eea
where $\mathcal{I}^k_{N_L }=\mathcal{I}^{k\dag}_{N_R},
\mathcal{I}_{D_R}^k= \mathcal{I}_{D_L }^{k\dag}$ and $k$ labels the
 $\phi_S,\phi_S^*$ flavons. Notice that in \eq{4comp} and for the rest of the paper we will indicate with $N_i$ the four RH neutrinos of the model under study, while in sec. 2 $N_i$ was referred to the group representations.

First of all we   change  basis going in the basis in which   $ \mathcal{M}^>_N$ is diagonal.  Thus we have
\bea
\label{def}
\hat{\mathcal{M}}^>_N&=&U_R^T \cdot  \mathcal{M}^>_N \cdot U_R= \mbox{Diag} (M_{\nu T},M_{\nu T},M_{\nu T},M_{\nu 4})=
\mbox{Diag} \,(M_{{N_i}_{(i=1,2,3)}},M_{N_4})\,,\nn\\
&&\nn\\
U_R&=& \frac{1}{\sqrt{2}} \left( \begin{array}{cccc}\sqrt{2}&0&0&0\\ 0&1 &-i&0\\0&1& i&0\\0&0&0 &\sqrt{2}\end{array}\right)\,,\nn\\
&&\nn\\
\hat{Y}^>_D&=& U_R^\dag \cdot Y^>_D =\frac{y_1}{2\sqrt{2}} \left( \begin{array}{ccc}\sqrt{2}&0&0\\ 0&1 &1\\0&-i& i\\0&0&0 \end{array}\right)\,.
\eea
It proves useful to write the interaction matrices  as
\bea
\label{Idef}
&& 	\mathcal{I}^k_{N_R }(\phi^*_{S_k})_{ij} =({\mathcal{I}}^>_{N_R})^{k}_{ij}\, {\phi}_{k}\,,  \nn\\
&&\nn\\
&&\mathcal{I}_{D_R }^{k}(\phi_{S_k}^*)_{ij} ={(\mathcal{I}}^>_{D_R})^{k}_{ij} \,{\phi}_{k}
\eea
and similarly for $\mathcal{I}^k_{N_L}(\phi_{S_k})_{ij},\mathcal{I}_{D_L }^{k}(\phi_{S_k})_{ij}$.
The ${\mathcal{I}}_{D_R}^{k}$ are $3\times 4$ matrices, being 4 the
total number of RH neutrinos. Equation \eqref{Idef} may appear as
redundant since in our model $\phi_S$ couples always linearly. However
the advantage of our notation is that it holds even when operators of
dimension higher than 5 are included.

\subsection{CP asymmetries}
\label{sec:cpa}
\begin{figure}
  \centering
  \includegraphics[width=8cm,height=3.4cm]{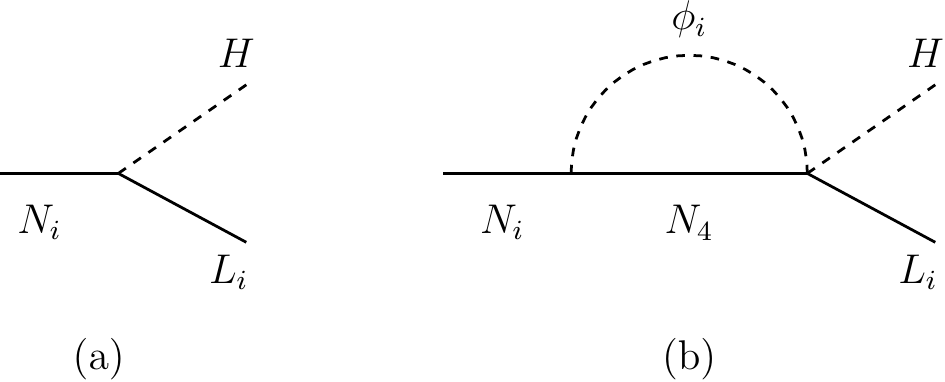}
  \caption{\it Tree-level and one-loop correction diagrams accounting for
    $\epsilon_{N_i}$.}
  \label{fig:cpasymm-diagrams}
\end{figure}

In the standard leptogenesis scenario the lightest RH neutrino CP
asymmetry, $\epsilon_N$, arises from the interference between the
tree-level decay Feynman diagram and the one-loop vertex and wave
function corrections \cite{Covi:1996wh}. Since $N_4$ does not have
renormalizable couplings to the lepton doublets such diagrams do not
exist in the case under consideration, regardless of the RH neutrino
mass spectrum.  New contributions due to the presence of the flavons
degrees of freedom exist and depend upon the RH neutrino spectrum:
\begin{enumerate}
\item
  \label{case1}
  \underline{{\bf The $\pmb{M_{N_i}>M_{N_4}}$ case}}: $N_i$ has
  standard $L_iH$ tree-level decays and, given the interactions in the
  Lagrangian \eqref{lag}, the only possible correction to this process
  arise from the one-loop correction to the effective vertex
  $N_i\phi_i L_i H$, as shown in
  fig. \ref{fig:cpasymm-diagrams}. Thus, the CP asymmetry in this case
  is obtained from the interference between diagrams
  \ref{fig:cpasymm-diagrams}(a) and \ref{fig:cpasymm-diagrams}(b).
\item
  \label{case2}
  \underline{{\bf The $\pmb{M_{N_4}>M_{N_i}}$ case}}: Since $N_4$
  couples to $N_i\phi_i$ at the renormalizable level a CP asymmetry
  for the two body decay process $N_4\to N_i\phi_i$ can be calculated
  from the interference between the corresponding tree-level diagram
  and the two-loop level diagram involving both effective couplings
  $N_4 L_i\phi_i H$ and $N_i L_i\phi_i H$ (since $N_4$ does not
  couples to lepton doublets at the renormalizable level the one-loop
  correction to the process $N_4\to N_i\phi_i$ does not exist). There
  is another option involving $N_4$ three-body decays induced by the
  effective coupling $N_4 L_i\phi_i H$. In this scenario a one-loop
  correction to the effective process does not exist either and the
  calculation of the CP asymmetry relies again on the two-loop level
  correction of the previous case.
\end{enumerate}

In case \ref{case1} the CP asymmetry arises in a different way
compared to the standard case but in what regards the generation of
the $B-L$ asymmetry there is no difference. In contrast, the cases in
\ref{case2} are quite different: for the three-body decay scenario the
differences are obvious, for the other scenario leptogenesis will take
place in two stages, a first stage in which an asymmetry in $N_i$ is
generated via the decays $N_4\to N_i \phi_i$ and a second stage in
which the asymmetry in $N_i$ is partially transfered to the lepton
doublets via $N_i$ decays and scatterings (a scenario of this kind has
been discussed in
\cite{AristizabalSierra:2007ur,JosseMichaux:2011ba}).  Note that in
this case the CP asymmetry, being a two-loop order effect, would most
likely yield a very tiny $B-L$ asymmetry.  All these scenarios however
exhibit a common feature, the generation of a $B-L$ asymmetry takes
place in the flavor symmetric phase. So from now on we will focus on
case \ref{case1}, that as was already pointed out resembles standard
leptogenesis.

\begin{figure}
  \centering
  \includegraphics[width=11cm,height=7cm]{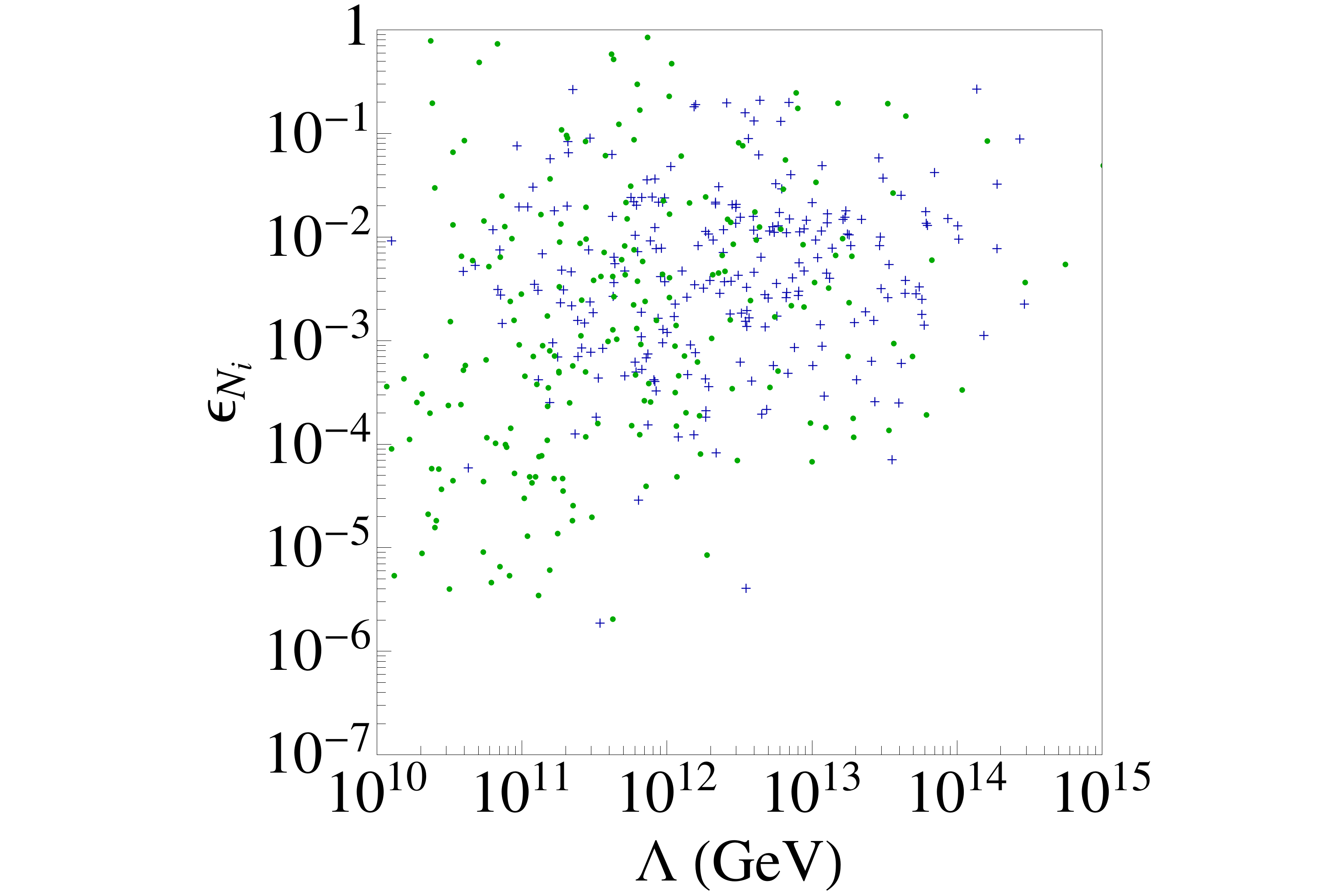}
  \caption{\it $\epsilon_{N_i}$ as a function of the cutoff scale
    $\Lambda$ for the spectra shown in fig. \ref{fig:mbb}. Green
    points (blue crosses) corresponds to QD-NO and NH (QD-IO and IH)
    neutrino mass spectra. The input parameters are generated as in
    fig.\ref{fig:mbb}. $\epsilon_{N_i}$ is computed according to
    \eq{eq:cp-asym} and \eq{anepsilon}.}
  \label{fig:CPasym}
\end{figure}

The CP asymmetry in the decay of $N_i$ is defined according to
\begin{equation}
  \label{eq:cp-asym}
  \epsilon_{N_i}=\sum_{k=e,\mu,\tau} \epsilon_{N_i}^{L_k}=
  \sum_{k=e,\mu,\tau}\frac{\Gamma_{N_i}^k - \bar \Gamma_{N_i}^k}
  {\Gamma_{N_i}^k + \bar \Gamma_{N_i}^k}\,,
\end{equation}
where $\Gamma_{N_i}^{L_k}$ ($\bar \Gamma_{N_i}^{L_k}$) denotes the
$N_i$ partial decay width for final states of flavor $k$ and carrying
$+1$ ($-1$) unit of lepton number, and $\epsilon_{N_i}^{L_k}$ are the
flavored CP asymmetries.  However, here we are working in the context
of an exact flavor symmetry. Since flavor is unbroken only flavor
conserving processes may happen, that in our framework means
$k=i$. Moreover flavor invariance and the representation used for our
right and left-handed neutrinos imply that the three RH neutrinos
produce the same amount of CP asymmetry and have exactly the same
dynamics.  Due to the complex  nature of the scalar field components $\phi_i$  running in the loop
\ref{fig:cpasymm-diagrams}(b) there is 
only one possible one-loop diagram of that type
(contrary to  the standard leptogenesis case
for the wave-function correction), so the interference between the
tree and one-loop level amplitudes (${\cal M}_0$ and ${\cal M}_1$)
involves only one term. For two-body decays this interference is
phase-space independent and consequently the calculation of
$\epsilon_{N_i}$ can be simply done in terms of the products of ${\cal
  M}_0$ and ${\cal M}_1$ and approximating the denominator of
\eqref{eq:cp-asym} with $|{\cal M}_0|^2$
\cite{AristizabalSierra:2007ur}. In the limit $M_{N_i}>>M_{N_4},
M_\phi$ the CP asymmetry can be written as
\begin{equation}
\label{anepsilon}
  \epsilon_{N_i}^{\ell_i}=-\frac{1}{8\pi}\frac{1}{|Y^>Y^{>\dagger}|^2_{ii}}
  \frac{M_{N_i}}{\Lambda}\mathfrak{I}\mbox{m}
  \left[
    (Y^>\,\mathcal{I}^k_{D_R})_{i4}
  \right]\,.
\end{equation}
In the general case, without assuming any large hierarchy among the
heaviest RH neutrino representation, the lightest one and the flavons,
the expression for $\epsilon_{N_i}$ is far more
involved. Fig. \ref{fig:CPasym}, obtained from the exact expression,
shows the possible values of the CP asymmetry as a function of the
effective cut-off scale $\Lambda$.

\subsection{Generation of the $B-L$ asymmetry}
\label{sec:bmla}
The generation of the $B-L$ asymmetry is entirely determined by $N_i$
dynamics but its final value depends on the washout induced by the
$A_4$ flavor singlet, $N_4$. Thus, leptogenesis in the case we are
interested in is a two-step process: generation of the $B-L$ asymmetry
and its subsequent washout via $N_4$ interactions (such scenario has
been analysed in the context of type-III seesaw in
\cite{AristizabalSierra:2010mv}). In what follows we will analyze both
stages in the unflavored regimen.
\subsubsection{$\pmb{N_i}$ dynamics}
\label{sec:bml-gen}
The determination of the $B-L$ asymmetry relies on the solution of the
kinetic equations for the $N_i$ abundance and the $B-L$ asymmetry
itself. At leading order in the coupling $y_1$, that is to say
including only $N_i\to L H$ decays and inverse decays and $\Delta L=2$
scatterings ($L H^\dagger\leftrightarrow L H^\dagger$ and $\bar L
H\leftrightarrow \bar L H$) \footnote{The inclusion of these processes
  is mandatory to obtain kinetic equations with the correct
  thermodynamical behavior \cite{Giudice:2003jh,Nardi:2007jp}.}, they
can be written according to
\begin{align}
  \label{eq:BEQs1-i}
  \frac{dY_{N_i}}{dz_i}&=-\frac{1}{s(z_i)H(z_i)z_i}
  \left(\frac{Y_{N_i}(z_i)}{Y_{N_i}^\text{Eq}(z_i)} - 1\right)
  \gamma_{D_i}(z_i)\,,\\
  \label{eq:BEQs2-i}
  \frac{dY_{\Delta_{B-L}}^T}{dz_i}&=-\frac{1}{s(z_i)H(z_i)z_i}
  \left[
  \left(\frac{Y_{N_i}}{Y_{N_i}^\text{Eq}(z_i)} - 1\right)\epsilon_{N_i}
  +\frac{Y_{\Delta_{B-L}}}{2 Y_\ell^\text{Eq}}
  \right]
  \gamma_{D_i}(z_i)\,,
\end{align}
where $z_i=M_{N_i}/T$, $s$ is the entropy density, $Y_X=n_X/s$ (with
$n_X$ the $X$ number density), $Y_L=2Y_\ell + Y_e$ (the lepton
asymmetry distributed in left-handed and RH degrees of freedom),
$H(z)$ is the expansion rate of the Universe and the reaction density
$\gamma_{D_i}(z_i)$ is given by
\begin{equation}
  \label{eq:reaction-dens}
  \gamma_{D_i}(z_i)=\frac{1}{8\pi^3}\frac{M_{N_i}^5}{v^2}
  \frac{K_1(z_i)}{z_i}\tilde m_T\,,
\end{equation}
with $v\simeq174$ GeV, $K_1(z_i)$ the modified Bessel function of
first-type and the parameter $\tilde m_T=v^2|y_1|^2/M_{N_i}$. An exact
solution of the kinetic equations in (\ref{eq:BEQs1-i}) and
(\ref{eq:BEQs2-i}) can only be done numerically, however a reliable
approximate solution can be found \cite{Buchmuller:2004nz}, which we
now discuss in turn. Equations (\ref{eq:BEQs1-i}) and
(\ref{eq:BEQs2-i}) can be recasted according to
\begin{align}
  \frac{dY_{N_i}}{dz_i}&=-D_T(z_i)
  \left[
    Y_{N_i}(z_i) - Y_{N_i}^\text{Eq}(z_i)
  \right]\,,\nonumber\\
  \label{eq:BEQ-rew2}
  \frac{dY_{\Delta_{B-L}}^T}{dz_i}&=-\epsilon_{N_i}D_T(z_i)
    \left[
    Y_{N_i}(z_i) - Y_{N_i}^\text{Eq}(z_i)
  \right]
  -
  W^T_{ID}(z_i)Y_{\Delta_{B-L}}\,,
\end{align}
where the new decay and inverse-decay functions read
\begin{align}
  \label{eq:decayFun}
  D_T(z_i)=K_T \,z_i\,\frac{K_1(z_i)}{K_2(z_i)}
  \quad
  \mbox{and}
  \quad
  W^T_{ID}(z_i)=\frac{1}{4}K_T \,z_i^3\,K_1(z_i)\,,
\end{align}
with $K_T=\tilde m_T/m_\star$ ($m_\star = 8\pi v^2
H(z_i=1)/M_{N_i}^2=1.08\times10^{-3}$ eV) and $K_2(z_i)$ is the
modified Bessel function of the second-type. In terms of $K_T$ the
strong (weak) washout regimen is defined as $K_T\gg 1$ ($K_T\ll 1$).

The $B-L$ asymmetry is obtained from the formal integration of
eqs. (\ref{eq:BEQ-rew2}) by means of the integrating factor technique:
\begin{equation}
  \label{eq:bmlfromT}
  Y^T_{\Delta_{B-L}}(z_i)=-3\times\epsilon_{N_i}\,
  Y^\text{Eq}_{N_i}(z_i\to 0)\,\eta(z_i)\,.
\end{equation}
Here $\eta(z_i)$ is the efficiency function defined as \cite{Buchmuller:2004nz}
\begin{equation}
  \label{eq:eff}
  \eta(z_i)=-\frac{1}{Y^\text{Eq}_{N_i}(z_i\to 0)}
  \int_{z_0}^{z_i}\,dz'\,
  \frac{dY_{N_i}(z')}{dz'}
  e^{-\int_{z'}^{z_i}\,dz''\,W^T_{ID}(z'')}\,.
\end{equation}
Note that we have included a factor of 3 in (\ref{eq:bmlfromT}) to
account for the $N_i$ flavor degrees of freedom. The final $B-L$
asymmetry is therefore obtained for $z_i\to\infty$ once the parameters
$K_T$ and $\epsilon_{N_i}$ are specified.

\begin{figure}
  \centering
  \includegraphics[width=9cm,height=6cm]{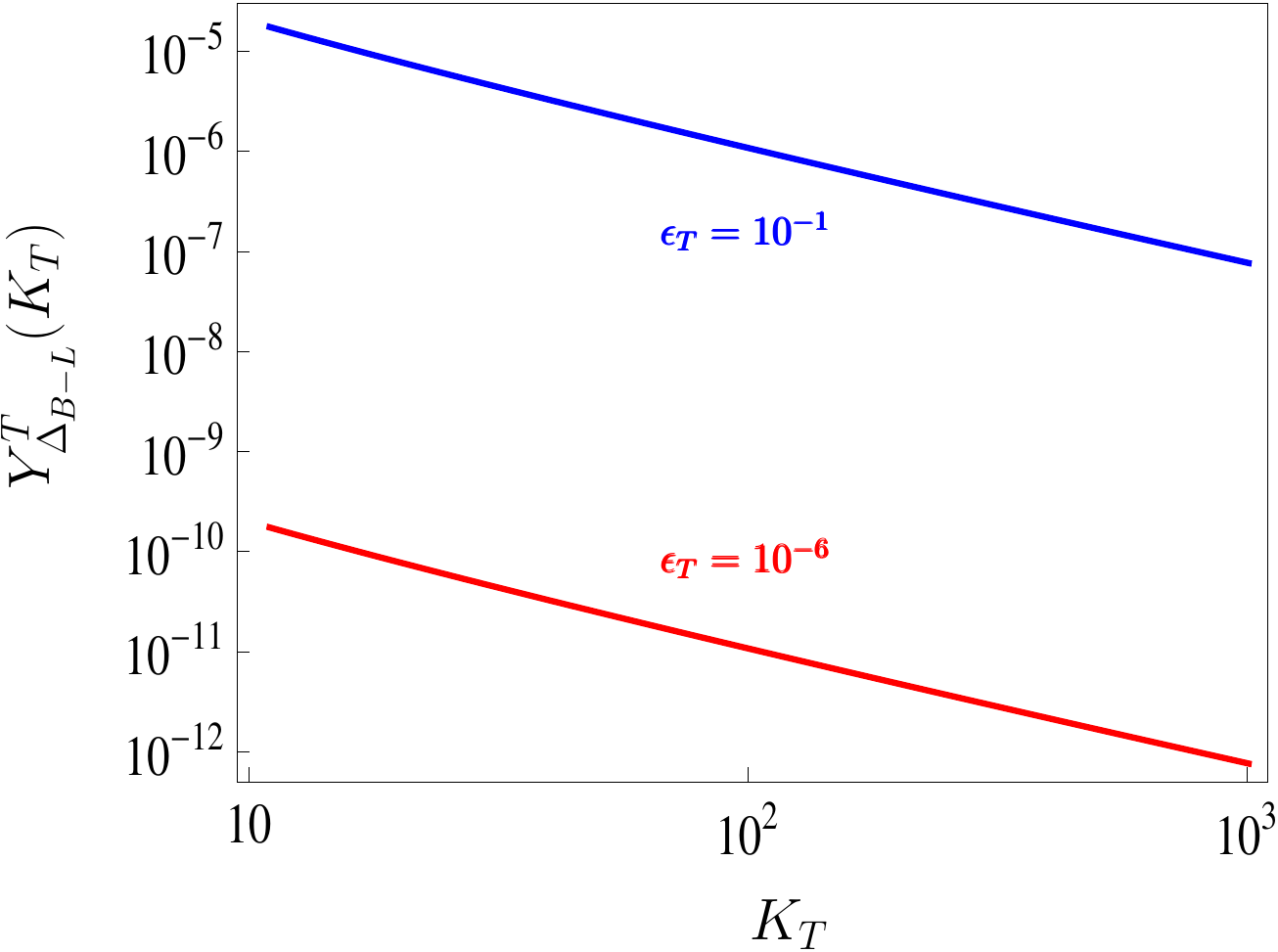}
  \caption{\it $B-L$ asymmetry produced by the $A_4$ flavor triplet
    ($N_i$) dynamics. The values $\epsilon_T=10^{-6}-10^{-1}$
    correspond to extreme cases.}
  \label{fig:ybmlvsKT}
\end{figure}
The problem of determining the final $Y^T_{\Delta_{B-L}}$ analytically
is thus {\it reduced} to find an approximate expression for the
efficiency function at $z_i\to\infty$ (efficiency factor). Such an
expression can be derived in the strong washout regimen by: $(i)$
noting that at low temperatures $Y_{N_i}(z_i)$ follows closely the
equilibrium distribution, so the replacement $dY_{N_i}(z_i)/dz_i\to
dY^\text{Eq}_{N_i}(z_i)/dz_i$ in eq.~(\ref{eq:eff}) can be done;
$(ii)$ replacing the washout function $W^T_{ID}(z_i)$ by $\overline
W_{ID}(z_i)=z_B\,W^T_{ID}(z_i)/z$, where $z_B$ is the minimum of the
function
\begin{equation}
  \label{eq:exp-fun}
  \psi(z',z_i)=-\ln\left(\frac{W^T_{ID}(z')}{z'}\right) + 
  \int_{z'}^{z_i}dz''\,W^T_{ID}(z'')\,.
\end{equation}
Following this procedure the efficiency factor can be derived
\cite{Buchmuller:2004nz}:
\begin{equation}
  \label{eq:eff-appr}
  \eta=\frac{2}{K_T\,z_B(K_T)}
  \left(
  1-e^{-K_T\,z_B(K_T)/2}  
  \right)\,,
\end{equation}
with
\begin{equation}
  \label{eq:zb}
  z_B(K_T)=\frac{1}{2}\ln
  \left\{
    \frac{\pi\,K_T^2}{1024}
    \left[
      \ln
      \left(
        \frac{3125\,\pi\,K_T^2}{1024}
      \right)
    \right]^5
  \right\}\,.
\end{equation}
With eqs. (\ref{eq:eff-appr}) and (\ref{eq:zb}) at hand we can
determine the maximum and minimum (still consistent with the measured
baryon asymmetry) $B-L$ asymmetry one can get through $N_i$
dynamics. The results are displayed in
figure~\ref{fig:ybmlvsKT}. Particularly relevant is the maximum value
$Y^T_{\Delta_{B-L}}\simeq 10^{-5}$ as it allows to derive an upper
bound on the washout induced by $N_4$ dynamics.
\subsubsection{$\pmb N_4$  washout}
\label{sec:washout}
\begin{figure}
  \centering
  \includegraphics[width=12cm,height=2cm]{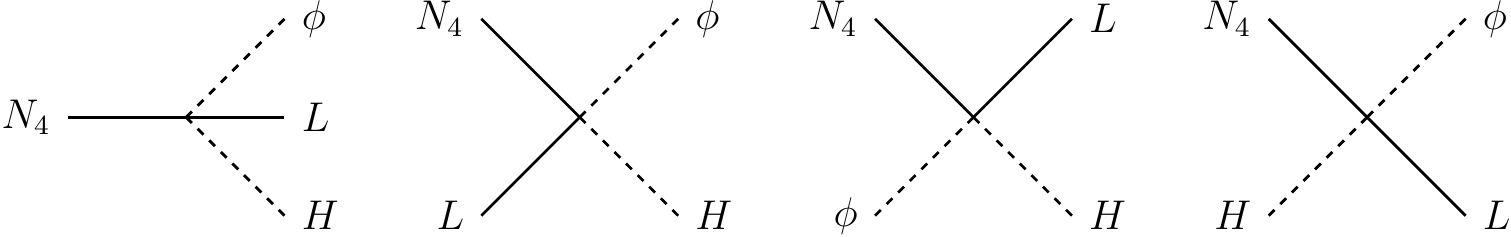}
  \caption{\it Relevant $1\leftrightarrow 3$ and $2\leftrightarrow 2$ $s$,
    $t$ and $u$ scattering processes accounting for the $A_4$ singlet
    washouts. }
  \label{fig:washout-process}
\end{figure}

The $B-L$ asymmetry produced at $z_i\sim 1$ remains frozen up to the
temperature at which $N_4$ washouts become effective,
$z_4=M_{N_4}/T\sim 1$. Since $N_4$ couples to lepton doublets via an
effective five-dimensional operator the dynamics of $N_4$ washouts, at
leading order in the couplings, involves not only the processes $N_4
\leftrightarrow L \phi H$ but the $2\leftrightarrow 2$ $s$, $t$ and
$u$ channel scatterings (see figure \ref{fig:washout-process}), in
contrast to the standard leptogenesis scenario. The derivation of the
corresponding kinetic equations in this case is tricky and requires
-even at leading order in the couplings- the inclusion of
$3\leftrightarrow 3$ and $2\leftrightarrow 4$ scattering processes
(see ref. \cite{AristizabalSierra:2009bh} for more details). Since
$\epsilon_{N_4}=0$ the kinetic equations accounting for the $N_4$
washouts can be written according to
\begin{align}
  \label{eq:washout-beqs1}
  \frac{dY_{N_4}}{dz_4}&=-\frac{1}{s(z_4)H(z_4)z_4}
  \left(\frac{Y_{N_4}(z_4)}{Y_{N_4}^\text{Eq}(z_4)} - 1\right)
  \gamma_\text{tot}(z_4)\,,\\
  \label{eq:washout-beqs2}
  \frac{dY^S_{\Delta_{B-L}}}{dz_4}&=
  -\frac{Y^S_{\Delta_{B-L}}}{2 Y^{\text{Eq}}_\ell}
  \left[
    \gamma_\text{tot}(z_4)
    + \left(\frac{Y_{N_4}(z_4)}{Y_{N_4}^{\text{Eq}}(z_4)} - 1\right)
    \gamma^s_{2\to 2}(z_4)
  \right]\,.
\end{align}
where $\gamma^s_{2\to 2}(z_4)$ is the reaction density for the
$2\leftrightarrow 2$ $s$-channel scattering process and
$\gamma_\text{tot}(z_4)$ involves the reaction densities for the full
set of processes shown in figure \ref{fig:washout-process}, namely
\begin{equation}
  \label{eq:gamma-tot}
  \gamma_\text{tot}(z_4)=\gamma_{1\to 3}(z_4) + \sum_{C=s,t,u} 
  \gamma^C_{2\to 2}(z_4)\,.
\end{equation}
As explained in appendix \ref{sec:app-washout} all the reaction
densities can be written in terms of the total decay width
$\Gamma(N_4\to L H\phi)$, which we have calculated to be
\begin{equation}
  \label{eq:total-decay-width}
  \Gamma(N_4\to L H\phi)=\frac{1}{192\pi^3}
  \frac{M_{N_4}^3}{\Lambda^2} |y_3|^2\,.
\end{equation}
In terms of the reaction densities given in (\ref{reaction-dens-def})
the kinetic equations in (\ref{eq:washout-beqs1}) can be rewritten in
such a way they resemble eqs. (\ref{eq:BEQ-rew2}):
\begin{align}
  \label{eq:washout-rewritten}
  \frac{dY_{N_4}}{dz_4}&=-D_S(z_4)
  \left(\frac{Y_{N_4}(z_4)}{Y_{N_4}^\text{Eq}(z_4)} - 1\right)\,,
  \nonumber\\
  \frac{dY^S_{\Delta_{B-L}}}{dz_4}&=-W^S_{ID}(z_4)Y^S_{\Delta_{B-L}}\,,
\end{align}
where now the functions $D_S$ and $W^S_{ID}$ are given by
\begin{align}
  \label{eq:newDandWID-fun}
  D_S(z_4)&=\frac{1}{4g_\star}\,K_S\,z_4^3
  \left[
    K_1(z_4)+\frac{3}{2}
    \left(
      S_s(z_4) + S_t(z_4)
    \right)
  \right]\nonumber\\
  W^S_{ID}(z_4)&=\frac{1}{4}K_S\,z_4^3
  \left[
    K_1(z_4)+\frac{3}{2}
    \left(
      \frac{Y_{N_4}}{Y_{N_4}^\text{Eq}}\,S_s(z_4) + S_t(z_4)
    \right)
  \right]\,.
\end{align}
\begin{figure}
  \centering
  \includegraphics[width=7.5cm,height=6cm]{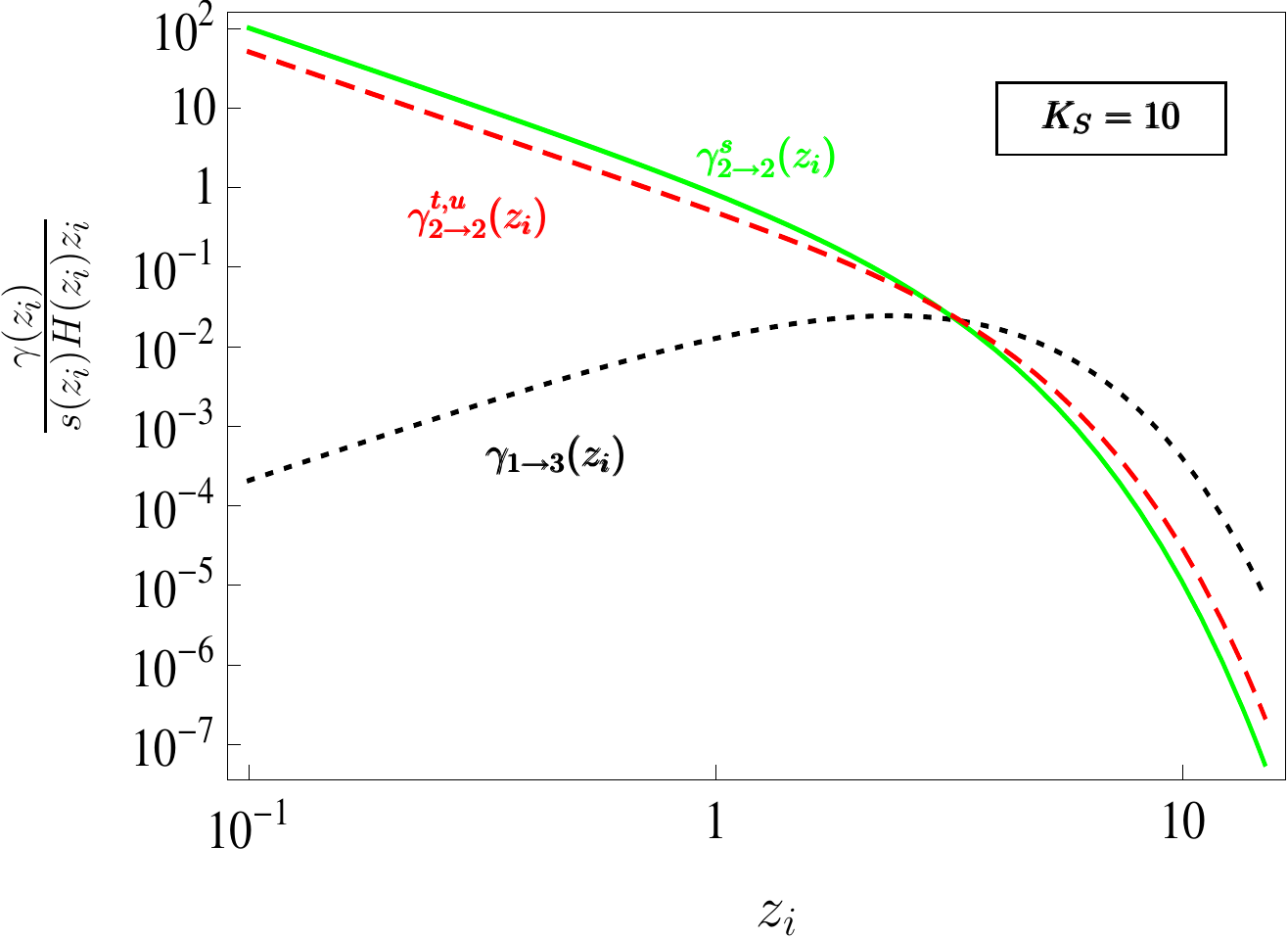}
  \includegraphics[width=7.5cm,height=6cm]{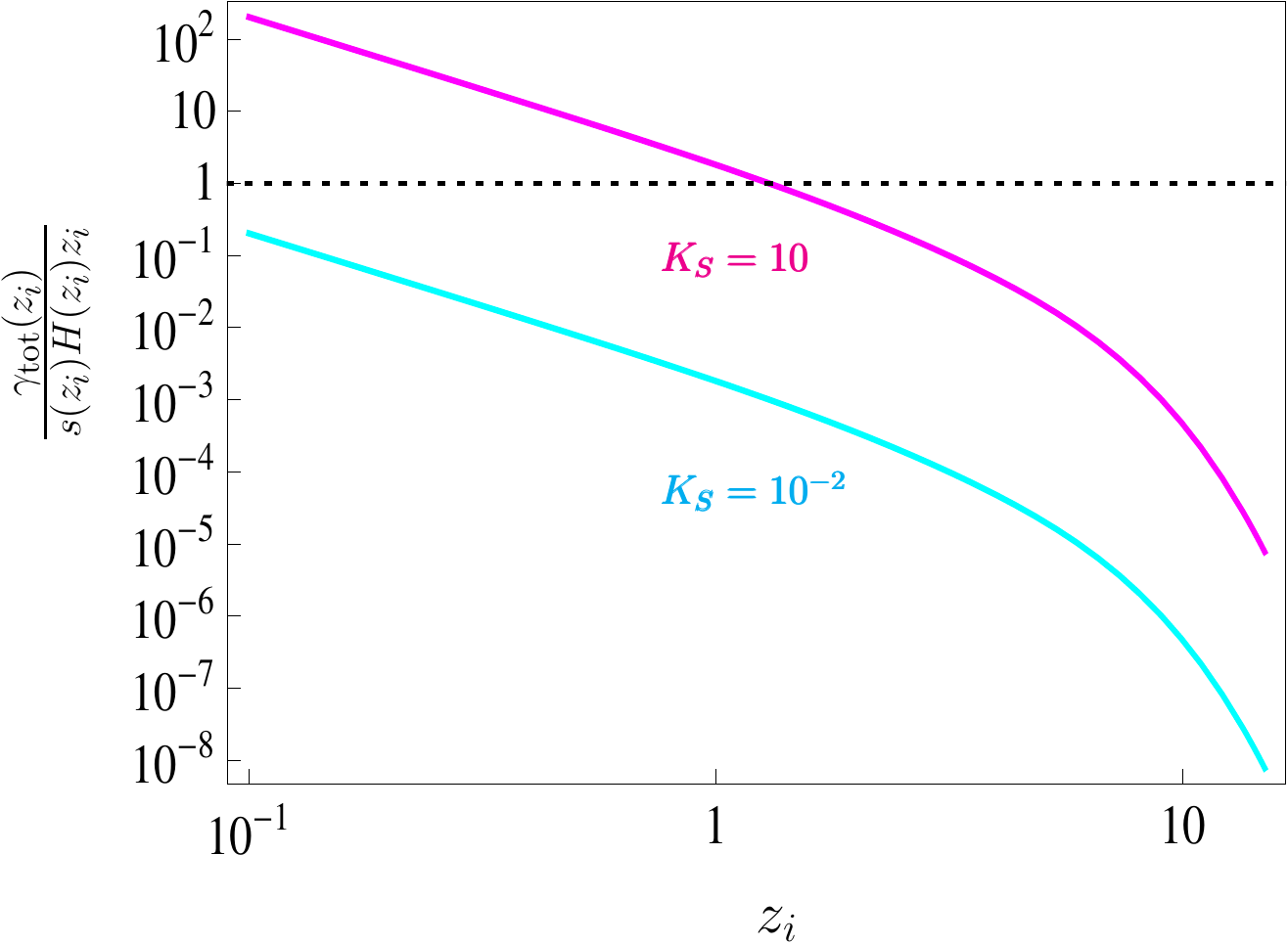}
  \caption{\it Reaction densities as a function of $z_i$ for the different
    processes present in $N_4$ washout (left panel) and total reaction
    densities for different values of the decay parameter $K_S$ (right
    panel). As in the standard case the strong washout (weak washout)
    regimen is defined according to $K_S\gg 1$ ($K_S\ll 1$).}
  \label{fig:reaction-dens}
\end{figure}
Some words are in order regarding these equations.  The relativistic
degrees of freedom are $g_\star=118$, as in our calculations we use
Maxwell-Boltzmann distributions, and the functions $S_{s,t}$ are given
in eqs. (\ref{eq:ssandstu-functions}) in the appendix. The decay
parameter $K_S$ is defined in the same way it is defined in the case
of $N_i$ dynamics, $K_S=\tilde m_S/m_\star$ but with
\begin{equation}
  \label{eq:ks}
  \tilde m_S=\frac{8\pi v^2}{M_{N_4}^2}\,\Gamma(N_4\to L H\phi)\,.
\end{equation}
The presence of the $2\to 2$ scattering processes may drive the system
to the strong washout regimen even when the $1\to 3$ process is
slow. Thus, the appropiate definition of the strong (weak) washout
regimen in this case reads:
\begin{equation}
  \label{eq:strong-washout-reg}
  \left.\frac{\gamma_\text{tot}(z_i)}{s(z_i)H(z_i)z_i}\right|_{z_i\sim 1}>1
  \quad(<1\; \mbox{for weak washout})\,.
\end{equation}
\begin{figure}
  \centering
  \includegraphics[width=7.5cm,height=6cm]{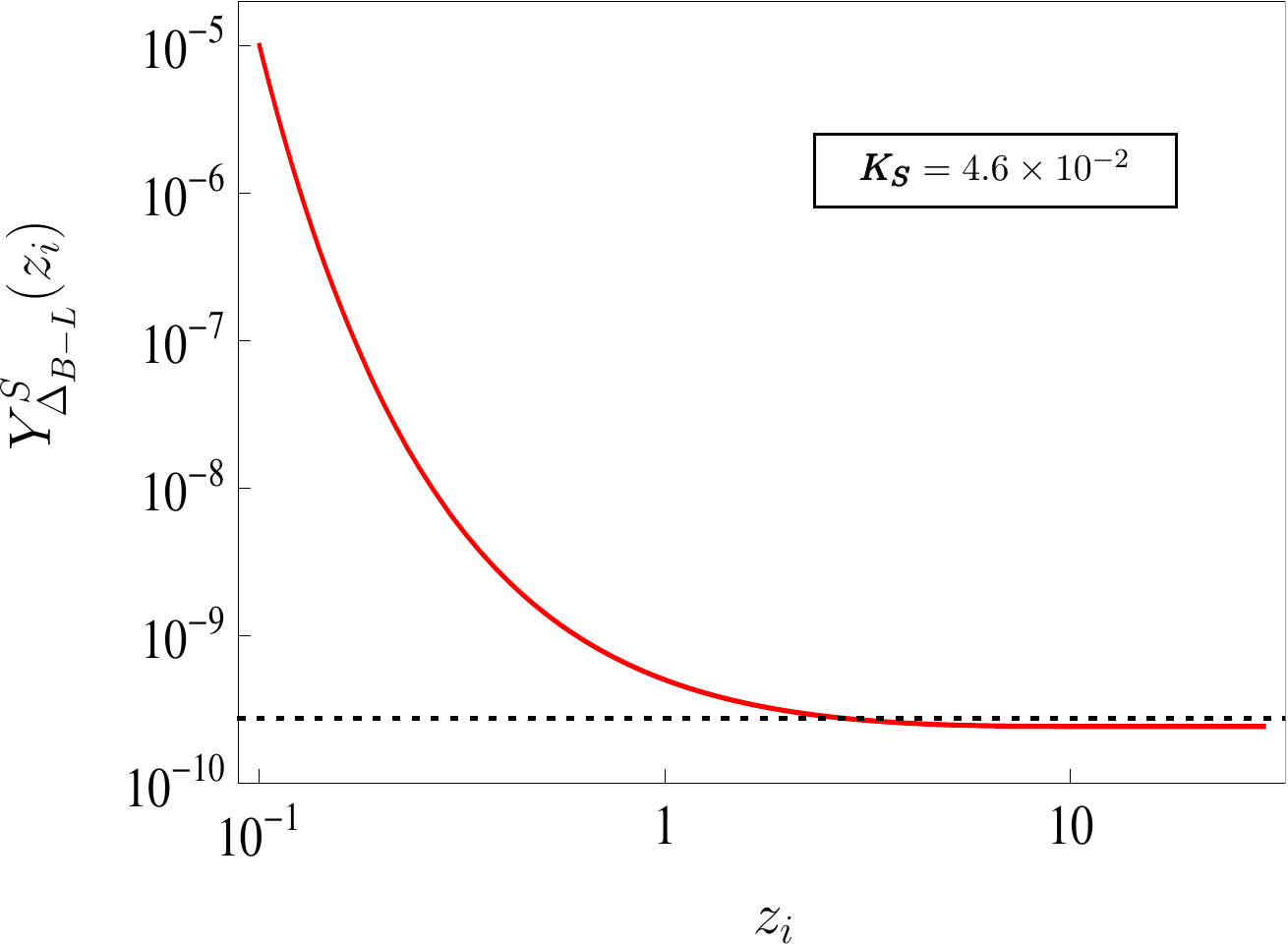}
  \includegraphics[width=7.5cm,height=6cm]{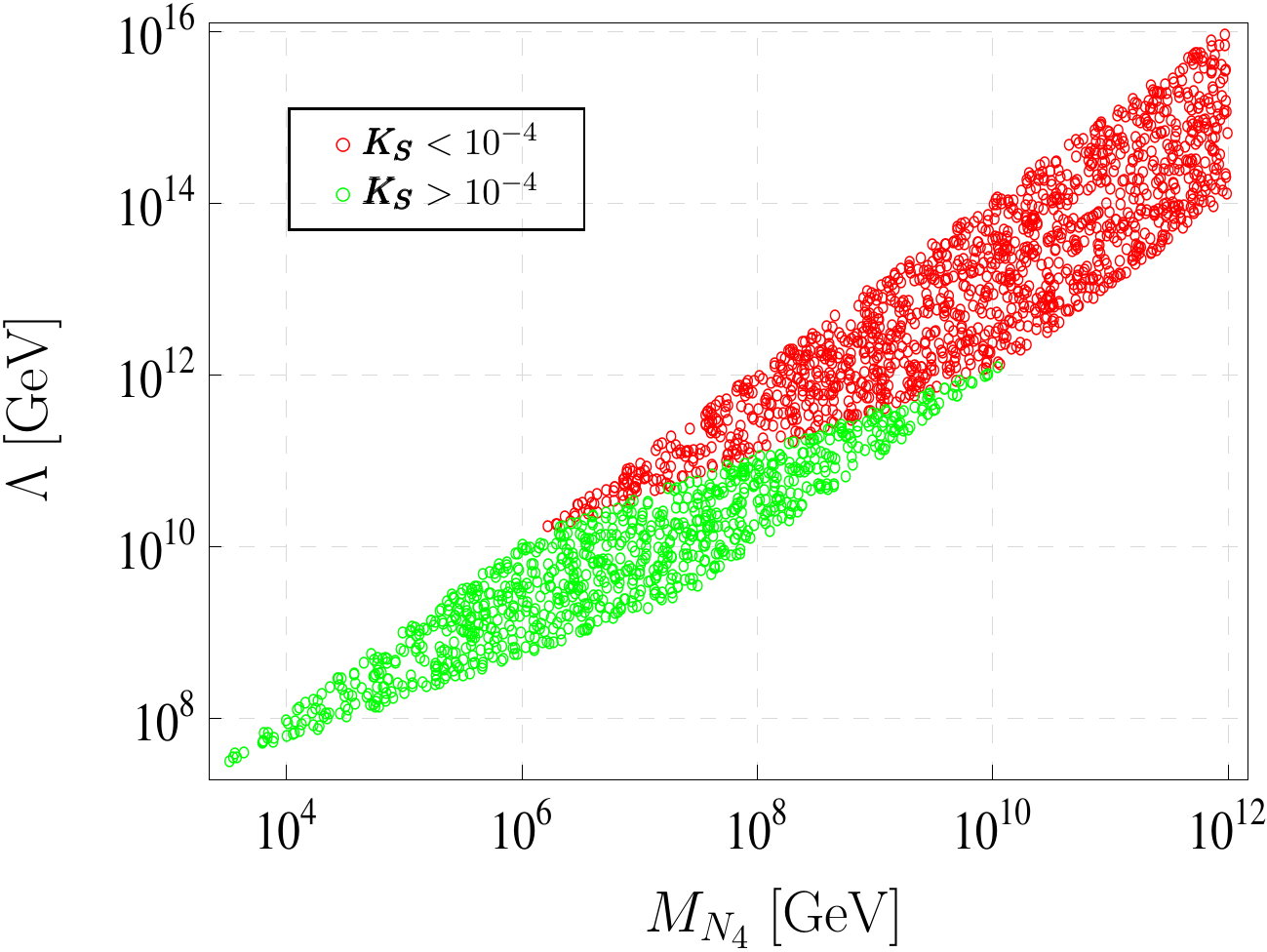}
  \caption{\it Washout induced by $N_4$ on the maximum $Y^T_{\Delta_{B-L}}$
    (see fig.~\ref{fig:ybmlvsKT}) generated in $N_T$ dynamics (left
    panel). The largest allowed $K_S$ for which the resulting
    $Y_{\Delta_B}$ still fits the measured value is
    $K_S^{\text{max}}=4.6\times 10^{-2}$, any value for which
    $K_S>K_S^{\text{max}}$ is excluded. On the right panel allowed
    regions of $\Lambda-M_{N_S}$ as required by the condition
    $K_S<K_S^{\text{max}}$.}
  \label{fig:MNlambdareg}
\end{figure}
Figure \ref{fig:reaction-dens} (left panel) shows an example in which
though $\gamma_{1\to 3}(z_i)/s(z_i)H(z_i)z_i|_{z_i\sim 1}\ll 1$ the
system is driven to the strong washout regimen by scattering
processes. Note however that the condition $K_S\gg 1$ ($K_S\ll 1$)
still determines the regimen in which the washout dynamics of $N_4$
takes place, as can be seen in figure \ref{fig:reaction-dens} (right
panel).

From the integration of eqs. (\ref{eq:washout-rewritten}) an upper
bound on $K_S$ for a given $Y^T_{\Delta_{B-L}}$ can be determined by
the condition of not erasing this asymmetry below $\sim 2.6\times
10^{-10}$. The maximum value $K_S^\text{max}$ is found for the largest
possible $B-L$ asymmetry generated in $N_i$ dynamics, that as has been
argued in sec. \ref{sec:bml-gen} we have found to be $\sim
10^{-5}$. Figure \ref{fig:MNlambdareg} (left panel) shows the final
$Y^S_{\Delta_{B-L}}$ matches the required value $\sim 2.6\times
10^{-10}$ (for $Y^T_{\Delta_{B-L}}=10^{-5}$) when $K_S^\text{max}\simeq
4.6\times 10^{-2}$, any value $K_S>K_S^\text{max}$ will induce a
washout that will damp the $B-L$ asymmetry below the allowed value.

Taking $|y_3|=10^{-2}$ the decay parameter $K_S$ becomes
\begin{equation}
  \label{eq:ksIeq10to-1}
  K_S=12\times 10^9 \,
  \left(\frac{M_{N_4}}{\text{GeV}}\right)\,
  \left(\frac{\text{GeV}}{\Lambda}\right)^2\,,
\end{equation}
with the purpose of placing the more stringent bounds on the
$\Lambda-M_{N_4}$ plane we fix $10^2<\Lambda/M_{N_4}<10^4$ and take
into account the restriction $K_S<K_S^\text{max}$. The results are
displayed in figure~\ref{fig:MNlambdareg} (right panel) where the
allowed $\Lambda-M_{N_4}$ region can be seen. Any discussion of
leptogenesis in the scenario we have considered here should be done at
least within that region.
\section{Conclusions}
\label{sec:conclusions}
In this paper we have study the necessary conditions that have to be
satisfied whenever the generation of the cosmic baryon asymmetry of
the Universe via leptogenesis takes place in the presence of a lepton
flavor symmetry accounting for lepton mixing. In the scenario we have
discussed, leptogenesis occurs in the flavor symmetric regime thus
before the flavons (that trigger the breaking of the flavor symmetry)
acquire vevs, accordingly the decays responsible for generating a net
$B-L$ asymmetry are liable of selection rules dictated by the flavor
symmetry.

In the core of the paper we exemplify how the general conditions for
the generation of the baryon asymmetry, in the flavor symmetric phase,
work by analysing a specific model based on the flavor symmetry $A_4$.
We briefly discussed the low energy phenomenology of the model and
studied in detail, by using the corresponding kinetics equations, the
generation of the baryon asymmetry. In the model considered, due to
the constraints imposed by $A_4$, the asymmetry proceeds through the
CP violating and out-of-equilibrium decays of the heaviest RH neutrino
$A_4$ representation. Subsequent washouts induced by the lightest
$A_4$ representation, being potentially dangerous, were properly taken
into account. Our onset shows these washouts can always be
circumvented and the correct amount of baryon asymmetry can be
produced.

In conclusion we have shown that under certain conditions, in models
containing flavor symmetries in the lepton sector, leptogenesis can
occur even in the flavor symmetric phase. The conditions we have
enumerated can be regarded as a general recipe for constructing lepton
flavor models in which the lepton number violating scale is above the
flavor symmetry breaking scale and the generation of the baryon
asymmetry proceeds via leptogenesis.
\section{Acknowledgments}
\label{sec:acknowl}
DAS would like to thank Luis Alfredo Munoz for discussions.  DAS is
supported by a FNRS belgian postdoctoral fellowship.
\appendix
\section{Appendix: Reaction densities for $1\to 3$ and $2\to 2$ processes}
\label{sec:app-washout}
In this appendix we present the relevant equations used in the
calculations discussed in section \ref{sec:washout}. The thermally
averaged reaction densities for $1\to 3$ and $2\to 2$ processes are
given by \cite{Giudice:2003jh}
\begin{align}
  \label{1to3and2to2}
  \gamma_{1\to3}&=n^\text{Eq}_{N_4}\,\frac{K_1(z_4)}{K_2(z_4)}\,\Gamma_{1\to 3}\,,\\
  \gamma^C_{2\to 2}&=\frac{M_{N_4}^4}{64\,\pi^5\,z_4}\,\int_1^\infty\,
  dx\,\sqrt{x}\,K_1({z_4\sqrt{x}})\,\widehat\sigma^C(x)\,,
\end{align}
where $x=s/M_{N_4}^2$ (with $s$ the center of mass energy) $C=s,t,u$,
$\Gamma_{1\to 3}\equiv\Gamma(N_4\to L H\phi)$ and
$\widehat\sigma(x)$, the reduced cross section, defined as
\begin{equation}
  \label{eq:reduced-cs}
  \widehat\sigma(x)=2\,M_{N_4}^2\,x\,\lambda(1,x^{-1},0)\,\sigma(x)
  \quad\mbox{with}\quad
  \lambda(a,b,c)=(a-b-c)^2-4bc\,.
\end{equation}
Neglecting the lepton doublets, Higgs and flavones masses we have found
for the differential cross sections the following results:
\begin{align}
  \label{eq:cross-sections}
  \frac{d\sigma^s}{dt}&=\frac{1}{16\pi}\,\frac{|y_3|^2}{\Lambda^2}\,
  \frac{1}{M_{N_4}^2}\frac{1}{1-x}\,,\nonumber\\
  \frac{d\sigma^t}{dt}&=\frac{1}{16\pi}\,\frac{|y_3|^2}{\Lambda^2}\,
  \frac{1}{M_{N_4}^2}\frac{1}{(1-x)^2}\,
  \left(1-\frac{t}{M_{N_4}^2}\right)\,,\nonumber\\
  \frac{d\sigma^u}{dt}&=\frac{1}{16\pi}\,\frac{|y_3|^2}{\Lambda^2}\,
  \frac{1}{M_{N_4}^2}\frac{1}{(1-x)^2}\left(x+\frac{t}{M_{N_4}^2}\right)\,.
\end{align}
Integrating over $t$ in the range $t_-=M_{N_4}^2(1-x)$ and $t_+=0$ and
using the definition for the reduced cross section,
eq. (\ref{eq:reduced-cs}), we get
\begin{align}
  \label{eq:reduced-cs-finalR}
  \widehat \sigma^s(x)&=\frac{|y_3|^2}{8\,\pi}
  \left(\frac{M_{N_4}}{\Lambda}\right)^2\,
  \frac{(x-1)^2}{x},\\
  \widehat \sigma^{t,u}(x)&=\frac{|y_3|^2}{16\,\pi}
  \left(\frac{M_{N_4}}{\Lambda}\right)^2\,
  \frac{(x^2-1)}{x}\,.
\end{align}
With these results at hand and taking into account the expression for
$\Gamma_{1\to3}$ given in
eq. (\ref{eq:total-decay-width}) the different reaction densities
become
\begin{align}
  \label{reaction-dens-def}
  \gamma_{1\to 3}&=\frac{M_{N_4}^3}{\pi^2}\,,
  \frac{K_1(z)}{z}\,\Gamma_{1\to 3}\,,\nonumber\\
  \gamma^s_{2\to 2}&=\frac{3M_{N_4}^3}{2\pi^2}\,
  \frac{S_s(z)}{z}\,\Gamma_{1\to 3}\,,\nonumber\\
  \gamma^{t,u}_{2\to 2}&=\frac{3M_{N_4}^3}{4\pi^2}\,
  \frac{S_{t,u}(z)}{z}\,\Gamma_{1\to 3}\,,
\end{align}
where the functions $S_{s,t,u}(z)$ are given by
\begin{equation}
  \label{eq:ssandstu-functions}
  S_s(z)=\int_1^\infty dx\frac{(x-1)^2}{\sqrt{x}}\,K_1(z\sqrt{x})
  \quad\mbox{and}\quad
  S_{t,u}(z)=\int_1^\infty dx\frac{x^2-1}{\sqrt{x}}\,K_1(z\sqrt{x})\,.
\end{equation}

\end{document}